\documentclass[]{aastex631}

\shorttitle{Evolving periodicity}
\shortauthors{Zheng et al.}

\begin{document}

\title{A period-increasing oscillation signal in a long gamma-ray burst}



\correspondingauthor{Da-Ming Wei; Stefano Covino}
\email{dmwei@pmo.ac.cn; stefano.covino@inaf.it}

\author[0000-0001-6076-9522]{Tian-Ci Zheng}
\affiliation{Key Laboratory of Dark Matter and Space Astronomy, Purple Mountain Observatory, Chinese Academy of Sciences, Nanjing 210023, China}
\affiliation{School of Astronomy and Space Science, University of Science and Technology of China, Hefei 230026, China}

\author[0000-0001-9078-5507]{Stefano Covino}
\affiliation{Osservatorio Astronomico di Brera, Istituto Nazionale di Astrofisica, Merate I-23807, Italy}
\affiliation{Como Lake centre for AstroPhysics (CLAP), DiSAT, Universit\`a dell'Insubria, Como I-22100, Italy}

\author[0000-0002-8385-7848]{Yun Wang}
\affiliation{Key Laboratory of Dark Matter and Space Astronomy, Purple Mountain Observatory, Chinese Academy of Sciences, Nanjing 210023, China}

\author[0000-0002-0584-8145]{Xiang-Dong Li}
\affiliation{School of Astronomy and Space Science, Nanjing University, Nanjing 210023, China}

\author[0000-0002-9758-5476]{Da-Ming Wei}
\affiliation{Key Laboratory of Dark Matter and Space Astronomy, Purple Mountain Observatory, Chinese Academy of Sciences, Nanjing 210023, China}
\affiliation{School of Astronomy and Space Science, University of Science and Technology of China, Hefei 230026, China}

\author[0000-0003-4977-9724]{Zhi-Ping Jin}
\affiliation{Key Laboratory of Dark Matter and Space Astronomy, Purple Mountain Observatory, Chinese Academy of Sciences, Nanjing 210023, China}
\affiliation{School of Astronomy and Space Science, University of Science and Technology of China, Hefei, 230026, China}



\begin{abstract}

Gamma-ray bursts (GRBs), the brightest electromagnetic bursts in the universe, are believed to originate from ultra-relativistic jets launched by the rapidly rotating central engine, either a disk-surrounded newly formed black hole (BH) or a magnetar.
Such a central engine potentially possesses rapidly evolving physical characteristics, as it is just born.
The caught of time-increasing frequency in the gravitational wave signals is the evidence of upon opinion.
Here we report a possible oscillatory signal identified in GRB\,131122B with periods increasing from 1.27 seconds to 4.02 seconds in a time interval of 16.75 seconds.
Such a peculiar oscillation signal has not been identified in GRBs before and its periodic evolution  could also be the quickest one found in the electromagnetic radiation window of astrophysics.
The precession of a misaligned accretion disk caused by the tidal disruption of a star by an intermediate-mass BH may be responsible for this signal.
This finding could open a new window to reveal the nature of the hiding central engine of GRBs.

\end{abstract}



\section{Introduction}
\label{sec:intro}
Quasi-periodic oscillations (QPOs) have been well observed in many different astrophysical sources such as X-ray binaries \citep{2019NewAR..8501524I}, soft gamma-ray repeaters \citep{2021Natur.600..621C} and so on.
As for cosmic gamma-ray bursts (GRBs; i.e, magnetar giant flares are excluded), QPOs are also predicted in some scenarios, including for instance the oscillations of the hypermassive neutron star formed in double neutron star mergers \citep{2005PhRvL..94t1101S}, the episodic accretion onto the central black hole (BH) \citep{2007ApJ...663..437M}, and the misalignment of the accretion disk formed in  mergers of stars (including also white dwarfs and neutron stars) with a BH \citep{1999ApJ...526..152F}.
Previous studies have reported some potential candidates with periods ranging from milliseconds to thousands of seconds in GRBs
\citep{2010AJ....140..224C,2016A&A...589A..98G,2021ApJ...921L...1Z,2023Natur.613..253C,2024ApJ...964..169Z}.
Motivated by the possible quick evolution of the period of the oscillation signal suggested in the numerical simulations \citep{2018MNRAS.474L..81L,2020arXiv200812381D} and by the oscillation frequency shift observations of some X-ray binaries \citep[XRBs;][]{2024ApJ...971..148Z} and in particular the gravitational wave (GW) events \citep{2016PhRvL.116f1102A,2017PhRvL.119p1101A}, in this work we carry out dedicated analysis of GRB sample on the time-evolving oscillation signature.
The most intriguing signal appears in GRB\,131122B, a long-duration GRB detected by \textit{Fermi}/GBM  and characterized by some distinct spikes.


\section{Methods}\label{sec11}

\subsection{Data selection and GRB\,131122B}\label{subsec:dataset}
Our goal is to identify GRBs exhibiting QPO signatures with periods that evolve over time.
An effective method to automatically identify potential QPO signals with time-evolving periodicity is currently unavailable.
Our initial sample was selected based on visual inspection of light curves recorded by various GRB detectors, i.e., the Fermi Gamma-ray Burst Monitor (GBM)\citep{2009ApJ...702..791M}, the Burst Alert Telescope (BAT)\citep{2005SSRv..120..143B} and the X-Ray Telescope (XRT) \citep{2005SSRv..120..165B} on the Neil Gehrels Swift Observatory, and the Compton Gamma Ray Observatory (CGRO) BATSE \citep{1994ApJS...92..229F}.
This screening method targets GRB light curves with at least four distinct spikes, pulses exhibiting similar patterns, and repetition times that follow a consistent increasing or decreasing trend.
We assume that the periodicity evolves in time according to a power-law as follow:
\begin{eqnarray}
\theta = {\rm A}(t - {\rm B} )^{\rm C} + {\rm D}.
\label{Eq:theta}
\end{eqnarray}
For a given light curve we selected the peak time of the $n$th pulse as $t_{\rm p,n}$ and assign the phase angle for the corresponding pulse as $\theta_n = 2n\pi$.
Therefore, the angular frequency over time is given by $\Omega = {\rm d}\theta /{\rm d} t = {\rm AC}(t - {\rm B} )^{\rm C-1}$, which
reveals a degeneracy in the specific scenario of $C\approx 0$.
For $\Omega \propto (t - {\rm B} )^{\rm -1}$, we employ a logarithmic (ln) function to fit the relationship between the phase angle and peak time
\begin{eqnarray}
\theta  = {\rm A}{\rm ln}(t-{\rm B})+C. 
\label{Eq:ln_fuc}
\end{eqnarray}
Two quantitative characteristic deviations of the phase angle are adopted to judge whether a signal is characterized by time-evolving periodicity, including
\begin{eqnarray}
|\theta_n - \theta(t_{\rm p,n})|< \frac{\pi}{4}, \nonumber\\
\displaystyle\sum^{n_{\rm max}}_{i=1}| \theta_n - \theta(t_{\rm p,n})| < n_{\rm max}^{1/2} \frac{\pi}{4},
\label{Eq:criterion3}
\end{eqnarray}
where $\frac{\pi}{4}$ and $n_{\rm max}^{1/2} \frac{\pi}{4}$ are two quantitative values, and $\theta(t_{\rm p,n})$ is determined by the best fit between the phase angle and the peak time of each pulse.

\begin{figure}[ht]
\centering
\includegraphics[angle=0,scale=0.5]{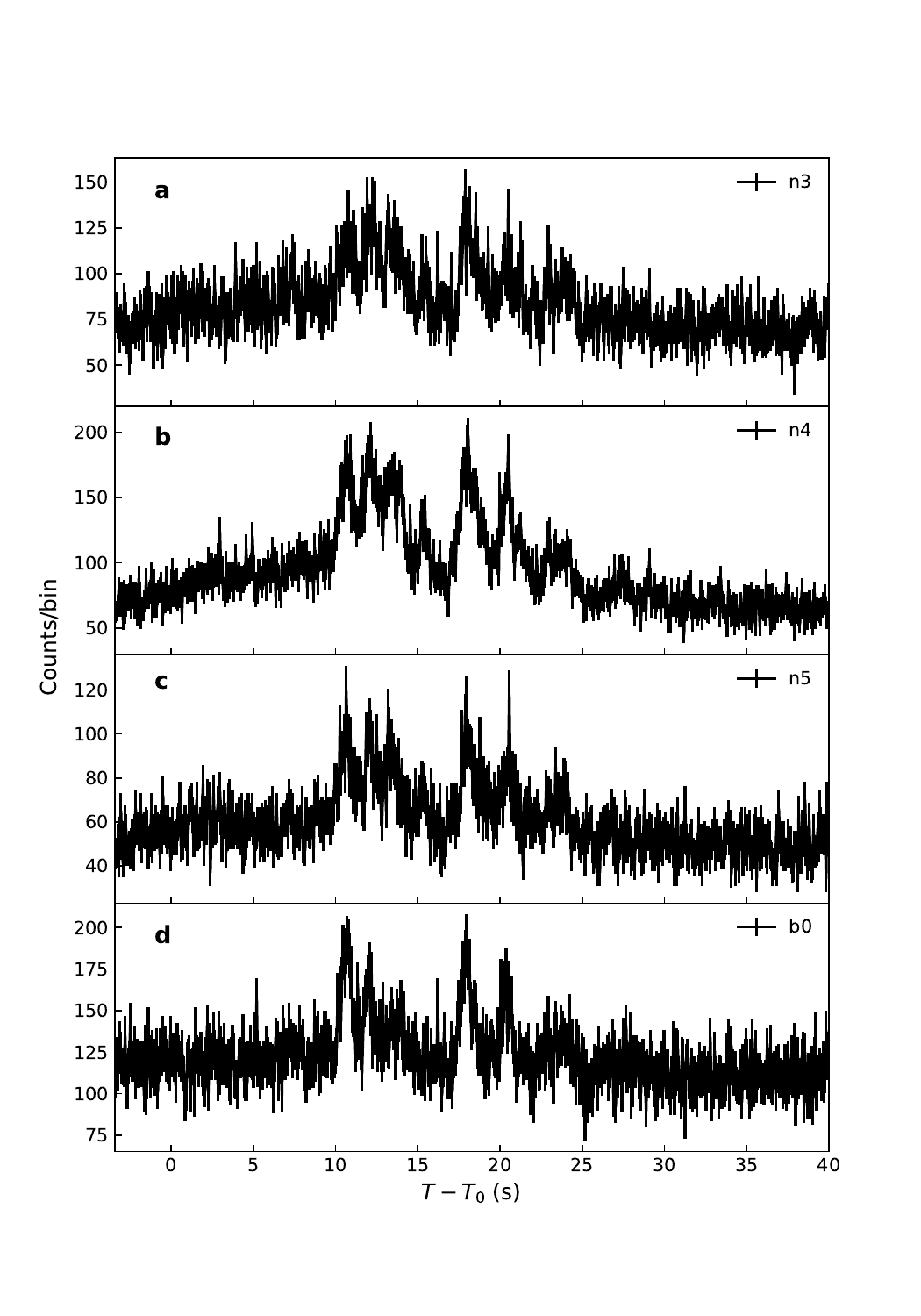}
\caption{\bf The light curves of GRB\,131122B recorded by individual detectors\rm. The emission has been well measured by three NaI detectors and one BGO detector, which are labeled as n3, n4, n5, and b0 in the four panels.}
\label{fig:lightcurves}
\end{figure}

\begin{figure}[ht]
\centering
\includegraphics[angle=0,scale=0.6]{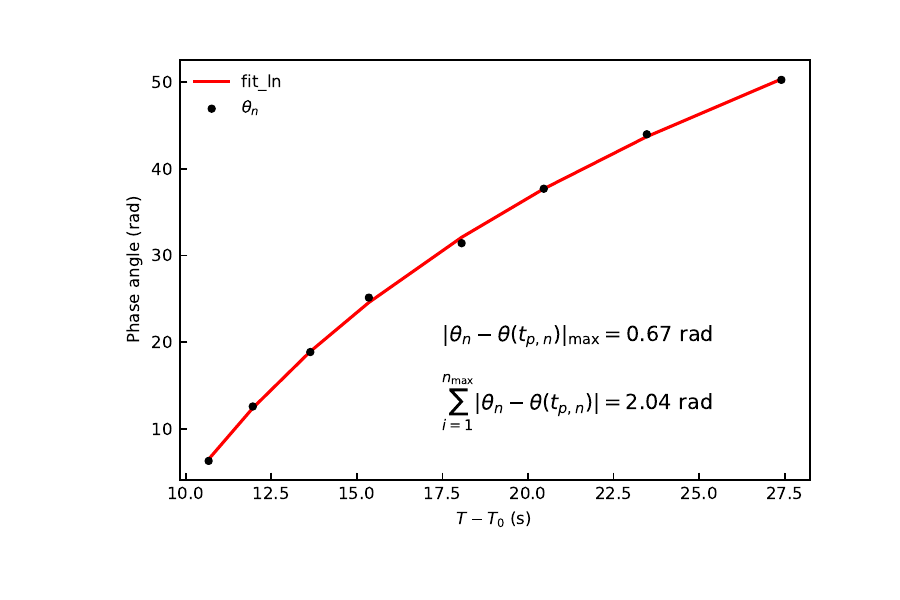}
\caption{\bf Fitting the phase angle and peak time relationship of GRB\,131122B\rm. }
\label{fig:cri_bn131122}
\end{figure}

The most interesting clue on the presence of a time-evolving oscillation signal appears in GRB\,131122B,
an event triggered \textit{Fermi}/GBM on 22 November 2013 at 11:45:05 UT (denote $T_0$ hereafter) and
was located at RA(J2000)  = 261.7$^{\circ}$, Dec(J2000) =  33.4$^{\circ}$  with an uncertainty of $1.7^{\circ}$. The duration ($T_{90}$) of this burst is reported to be $23.04 \pm 0.81$ seconds \citep{2020ApJ...893...46V}.
The initial GCN notices \footnote{https://gcn.gsfc.nasa.gov/other/406813508.fermi} for the event indicated a 90\% probability of it being a GRB, likely because the automated analysis was based on the first 4 seconds, whereas the main burst, consisting of eight pulses, began at about 10.0 seconds. This may have contributed to the absence of multi-band follow-up observations.
The \textit{Fermi}/GBM includes two types of detectors, 12 sodium iodide (NaI) detectors and 2 bismuth germanate (BGO) detectors.
The selecting Time Tagged Event (TTE) data from three NaI detectors (n3, n4, and n5) and one BGO detector (b0) were used to construct the multi-detector light curves for GRB\,131122B (Fig. \ref{fig:lightcurves}).
The light curves are binned with a time resolution of 64 ms, characterized by a
multi-pulse structure.
Among the NaI detectors, n4 has the lowest viewing angle for this GRB ($\alpha_{\rm n4} = 19.19^{\circ}$), which is much smaller than others ($\alpha_{\rm n3} = 65.25^{\circ}$ and $\alpha_{\rm n5} = 63.16^{\circ}$).
That is why its signal-to-noise ratio is the highest and in this work we focus on this set of data (note that for the b0 detector, the background is much higher than the NaI detectors). For this event we have $|\theta_n - \theta(t_{\rm p,n})|_{\rm max}  = 0.67 ~ {\rm rad}$ and $\displaystyle\sum^{n_{\rm max}}_{i=1}| \theta_n - \theta(t_{\rm p,n})|  = 2.04 ~ {\rm rad}$, well satisfying the constraints of Eq.(\ref{Eq:criterion3}).
The fitting result is shown in Fig \ref{fig:cri_bn131122}.

\subsection{Filter }\label{filter}
\begin{figure}[ht]
\centering
\includegraphics[angle=0,scale=0.8]{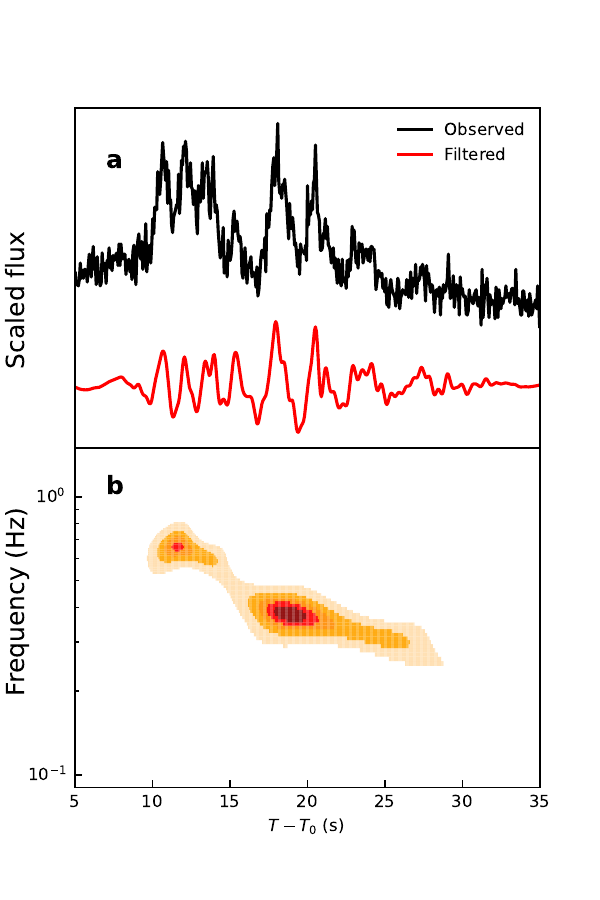}
\caption{\bf Bandpass filter result and the 2D plane contour plot of the WWZ power\rm.
{\bf a\rm}, The observed light curve of GRB\,131122B is in black, while the filtered light curve is in red.
The cutoff frequencies of the bandpass filter are set at 0.2 Hz and 2.0 Hz, respectively.
{\bf b\rm}, The 2D plane contour plot of the WWZ power of the filtered light curve and a time-declining oscillation signal appears. The deeper the colors, the higher the power values.
}
\label{fig:filter}
\end{figure}

The presence of background noise can lead to false signals appearing in the periodogram.
In the processing of GW data, background subtraction and filtering techniques are employed to recover the true evolutionary form of GW signals \citep{2017PhRvL.119p1101A}.
Here we implemented a band-pass filtering operation on the data, using two characteristic filter frequencies of 0.2 Hz and 2.0 Hz.
A Tukey window model is employed in the filter procedure.
The resulting filtered light curve is presented in Fig. \ref{fig:filter}a, while the corresponding spectrogram given by weighted wavelet Z-transform (WWZ)\citep{1996AJ....112.1709F} is shown in Fig. \ref{fig:filter}b.
The analysis reveals eight distinct pulses within the filtered light curve occurring between $T_0 + 10$ s to $T_0 + 29$ s.
These pulses are characterized by increasing time intervals, which correspond to decreasing frequency in the spectrogram.

\subsection{Light curve scaling and the frequency-decreasing oscillation signal}\label{subsec:normalization}

\begin{figure}[ht]
\centering
\includegraphics[angle=0,scale=0.6]{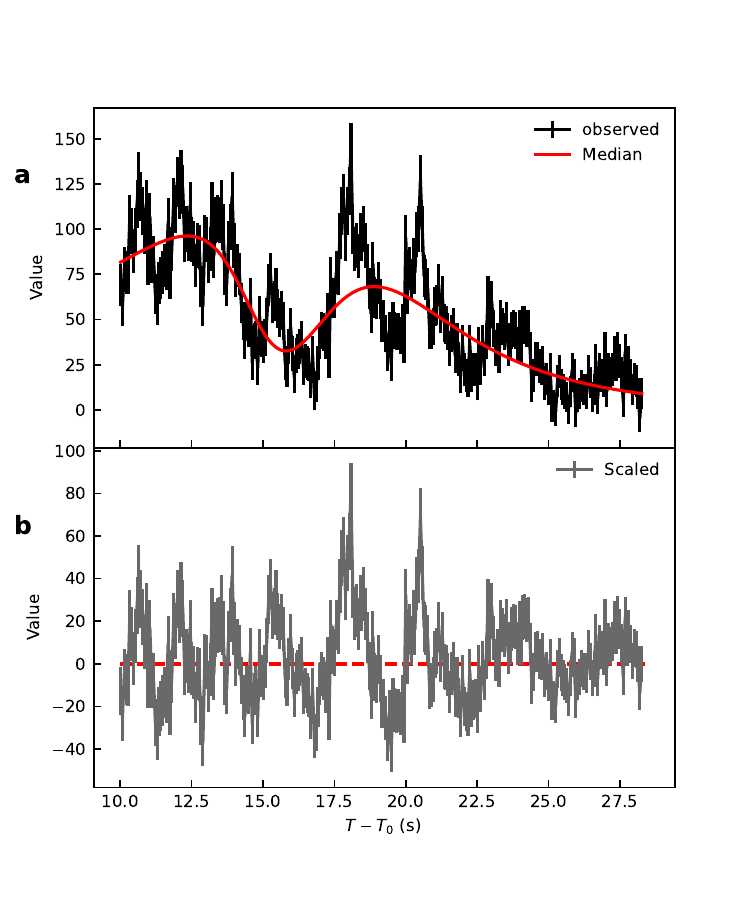}
\caption{\bf Scaling the observed light curve\rm.
{\bf a}, The fitting of observed light curve with a two smooth-broken power law function. The light curve is recorded by n4 detectors with energy covering 8 -900 keV. The data adopted in the fit covers a time range from 10.0 seconds to 28.3 seconds. A red line represents the fitting result, labeled as ``Median''.
{\bf b}, The observed light curve subtracts the ``Median" resulting in the so-called ``Scaled" light curve.
}
\label{fig:scaling}
\end{figure}

In this section, a scaling procedure is employed to analyze the periodicity in detail.
Motivated by the finding made in the data selection and filtering approach, we take a two smooth-broken power law function to fit the ``Median" of the light curve that resembles the possible low-frequency background component, i.e.,

\begin{equation}
f_b(t) = A_q\left(\left(\frac{t}{t_{\rm b,q}}\right)^{-a_{\rm 1,q} w}+\left(\frac{t}{t_{\rm b,q}}\right)^{-a_{\rm 2,q} w}\right)^{1/w} + A_h \left(\left(\frac{t}{t_{\rm b,h}}\right)^{-a_{\rm 1,h} w}+\left(\frac{t}{t_{\rm b,h}}\right)^{-a_{\rm 2,h}w}\right)^{1/w} ,
\label{Eq:2BPL}
\end{equation}
where the smooth index $w$ is set as $-1/3$, $A$ is the normalization constants, $a_1$  and $a_2$ represent the decay slopes before and after the break time $t_b$, respectively.
The subscripts $q$ and $h$ denote the first and second components, respectively.
The corner plot of the posterior parameters distribution of the fitting procedure is presented in Fig. \ref{figEXT:corner_2SMBPL}. Note that in our current fit,
$a_{\rm 1,h}$ is fixed as $-27$ to prevent the derived Median line from matching the rapid rise of the fifth pulse (leaving this parameter free will not change our result, as shown in Section \ref{secA3}). We then get the ``Scaled" light curve by subtracting the Median from the observation.
This procedure is demonstrated in Fig. \ref{fig:scaling}.
As shown in Fig. \ref{fig:scaling}b, the ``Scaled" light curve does not suffer from the contamination of the low-frequency component and is well suitable for further periodicity analysis. As mentioned above, the light curve of GRB\,131122B was rebinned into 64 ms, corresponding to a Nyquist frequency $1/0.064~ {\rm s} / 2 \sim 7.8 $ Hz. The observed eight pulses last about 20 seconds (from $T_0+10$ s to $T+30$ s), corresponding to the lowest frequency $1/20 = 0.05$ Hz.
We performed a wavelet transform on the Scaled data, utilizing a frequency range of $0.09 - 1.5$ Hz. The WWZ method-based spectrogram reveals a time-decreasing frequency (Fig. \ref{fig:t-e periodicity}c).
\rm

\subsection{Evolution form of periodicity}\label{subsec:evloution}
The time-evolving periodicity in GRB\,131122B has been suggested by both the increasing time intervals of eight pulses and the frequency-decreasing spectrogram of the filtered light curve.
The fitting of the relationship between the phase angle and peak time (see Section \ref{subsec:dataset}) suggests that the evolving angular frequency follows a form of $\Omega \propto t^{-1}$, indicating that the evolving phase angle ($\theta (t)$) can be expressed as $\theta (t) = \int \Omega (t) dt = N_\theta {\rm ln}(t - t_s) + \theta_s$.
In this context, we construct a function $V(t) = N_V {\rm cos}(\theta (t)) = N_V {\rm cos}(N_\theta{\rm ln}(t - t_s) + \theta_s)$ to fit the Scaled light curve.
Here, $t_s$ and $\theta_s$ represent the time and phase shifts respectively, $N_V$ signifies the average amplitude of eight pulses in the Scaled light curve, and $N_\theta$ is the normalization constant of the evolving phase angle.
By fitting the Scaled light curve with the function $V(t)$, we derive the following parameters: $N_V = 17.30^{+0.90}_{-0.87}$, $N_\theta = 39.64^{+0.92}_{-0.92}$, $t_s = 2.26^{+0.16}_{-0.17}$, and $\theta_s = 2.88^{+1.67}_{-1.72}$.
What needs illustration is that $\theta_s$ exhibits convergence within each $2\pi$ parameter space according to a corner plot of the posterior parameter distribution.
Therefore the present result is one of the feasible outcomes.
In this context, the real evolution of the phase angle has $\theta(t) =  39.64^{+0.92}_{-0.92}{\rm ln}(t-2.26^{+0.16}_{-0.17}) - 2.88^{+1.67}_{-1.72}$, with the corresponding angular frequency and frequency written as $\Omega (t) = 39.64^{+0.92}_{-0.92} (t - 2.26^{+0.16}_{-0.17})^{-1} $ and $f(t) = 6.31^{+0.15}_{-0.15}(t-2.26^{+0.16}_{-0.17})^{-1}$ respectively.
The curve $V(t)$ and evolving frequency $f(t)$ are depicted as the red dotted line in Fig. \ref{fig:t-e periodicity}b and Fig. \ref{fig:t-e periodicity}c, respectively.
Based on the derived $\theta (t)$, we estimate the real peak times ($t_{\theta = 2n\pi}$) of each pulse to occur at 10.63 s, 12.06 s, 13.72 s, 15.67 s, 17.95 s, 20.61 s, 23.73 s, and 27.38 s, with corresponding periods ($P_{\rm peak} = 2\pi/\Omega(t_{\theta = 2n\pi})$) of 1.27 s, 1.50 s, 1.78 s, 2.10 s, 2.47 s, 2.91 s, 3.42 s, and 4.02 s.
Thus, the observed QPO signature shows a progression of periods, increasing from 1.27 s to 4.02 s in a 16.75-second interval.

\subsection{Stretching Light curve and constant periodicity in phase space}\label{subsec:stretch}
In this section, we transform the Scaled light curve into phase space by substituting the derived relation $\theta (t)$ for $t$ in the light curve, i.e., $V(\theta) = 17.30 {\rm cos} \theta$, the resulting light curve is labeled as ``Stretched" (Fig. \ref{fig:t-e periodicity}d).
In phase space, the Stretched light curve can be modeled by a constant period and it is characterized by an uneven sampling, i.e., $\Delta \theta = \theta(t_{i+1}) - \theta(t_i)$ is not constant.
The analysis of the Stretched light curve using the WWZ method results in a horizontal spectrogram (see Fig. \ref{fig:t-e periodicity}e), and the analysis with  the Lomb-Scargle periodogram (LSP) shows a prominent peak in the periodogram ($P_{\rm LSP}$; Fig. \ref{fig:t-e periodicity}f). Both methods reveal a signal at the frequency $1/2\pi$ (/rad).
As we convert the time axis to the phase angle axis, the appropriate unit of frequency becomes ``/rad".

\subsection{Testing the evolving periodic component with the Gaussian process}\label{subsec:GP}
The Gaussian Process (GP) is a method that combines a deterministic component with a stochastic process, enabling direct analysis of the periodic component in an observed light curve \citep{2020ApJ...895..122C,2022ApJ...936...17H,2022MNRAS.513.2841C,2023ARA&A..61..329A,2025arXiv250105602G}.
GPs offer the possibility to model a time sequence with great flexibility in a probabilistic framework. A key role in any GP analysis is played by the kernel or covariance function applied to model the covariance possibly affecting the observed data. In this work we follow the framework discussed in \cite{2022ApJ...936...17H}. Noise affecting our data is described by a Damped Random Walk (DRW, i.e., an Ornstein-Uhlenbeck process in one dimension), whose kernel is simply a decaying exponential:
\begin{equation}
    k(|t_i - t_j|) = \sigma^2 \exp(-c_1 |t_i - t_j|)
\end{equation}
where $\sigma^2$ is the variance of the process and $c_1$ is the inverse of the decaying time-scale of the noise correlation. For the periodic component, we have employed a single exponentially decaying co-sinusoid:
\begin{equation}
    k(|t_i - t_j) = \sigma^2 \exp(-c_2 |t_i - t_j|) \cos (2\pi f |t_i - t_j|)))
\end{equation}
where $c_2$ is the inverse of the decaying time-scale of the QPO correlation, and $f$ represent the frequency of oscillation, respectively.

For an evolving periodicity, the frequency is obviously not constant but this is easily included in the GP analysis transforming the input data in the model by a stretching function as discussed above.
In this context, the phase of the stretching function is degenerate with the oscillation frequency, and both the reference time for the start of the oscillation and the amplitude of the oscillation in the GRB light curve are important:
\begin{eqnarray}
K_{\rm ep} &=& k_0 {\rm cos}(2\pi f|\theta(t_i) - \theta(t_j)|) \nonumber\\
&=& k_0 {\rm cos}(2\pi f |A {\rm ln}(t_i - B) + C - A {\rm ln}(t_j - B) - C|) \nonumber\\
&=& k_0 {\rm cos}(2\pi fA |{\rm ln}((t_i - B)/(t_j - B))|) \nonumber\\
&=& k_0 {\rm cos}(2\pi F \tau)),
\label{Eq:Ker_e_p}
\end{eqnarray}
where $k_0 = \sigma^2 \exp(-c_2 |t_i - t_j|)$, $F$ and $\tau$ are general oscillation frequency and time, respectively, $B$ represent a time shift.
We applied our GP model to the raw data collected from the n4 detector over the time range $T_0 + 10.0  - T_0 + 28.3 $ seconds and adopt a constant mean function equal to the average of the light-curve \citep{2025arXiv250105602G}.
We adopted large uniform logarithmic priors for all the scale parameters and the frequency. A truncated positive Normal distribution model the prior for the starting time of the oscillation. Priors are reported in Table\,\ref{tab:priors}.

\begin{table}[]
\centering
\begin{tabular}{c | c}
\hline
$\ln \sigma$ & $\mathbb{U} (-10,10)$ \\
$\ln c_1$ & $\mathbb{U} (-10,10)$ \\
$\ln c_2$ & $\mathbb{U} (-5,15)$ \\
$B$ & Truncated $\mathbb{N} (0,5)$ \\
$\ln F$ & $\mathbb{U} (\ln{0.05},\ln{8})$ \\
\hline
\end{tabular}
\caption{Priors adopted for the GP analysis. $\sigma$ is the variance of the kernel, $c_1$ and $c_2$ the length scale of the exponential and $F$ is the general oscillation frequency, and $B$ represents the time shift.}
\label{tab:priors}
\end{table}

Two models, one with noise only and another with the noise combined with an evolving periodic component are compared.
Based on the Bayesian inference, we derive the Bayes factor
${\rm ln} BF = {\rm ln}{\cal Z}_{\rm noise+QPO}-{\rm ln}{\cal Z}_{\rm noise} = 12.31$.
To address the significance of the proposed model vs a noise-only description we may need to correct this result for the trial factor deriving by the original sample size of $\sim 1000$ events with at least four peaks in their light curve.
This bring the significance of the time-evolving QPO component to slightly better than $2.8\,\sigma$. In addition, the choice of the starting and ending time for the analysis could also imply a correction, although not easy to quantify. The actual significance will therefore likely be somehow lower than the reported figure. On the other hand, this is the result of the analysis of one promising candidate among several more. If at least one of them should prove to be interesting at a level similar to GRB131122B their joint significance would be substantially higher. We leave the full analysis of these candidates to a future paper.

\section{Results}
\label{sect:result}

\begin{figure}[ht]
\centering
\includegraphics[angle=0,scale=0.5]{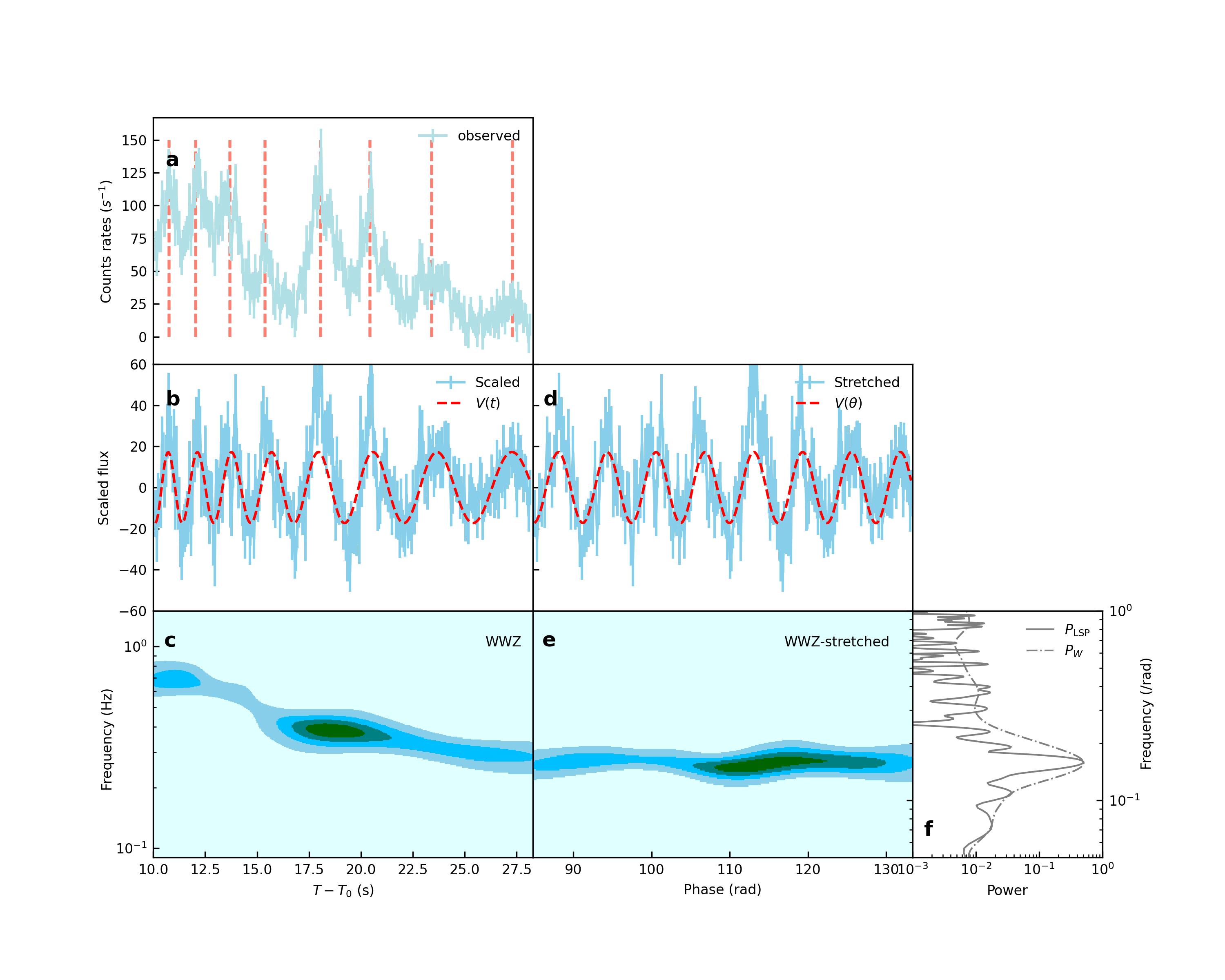}
\caption{{\bf The time-evolving periodicity of the oscillation signal of GRB\,131122B.}
{\bf a}, The eight spikes recorded by the n4 detector in the time interval of $10.0-28.3$ seconds after the trigger of the burst.
The pink dashed lines mark the estimated pulse peak times.
{\bf b}, The Scaled light curve and the
modeled light curve $V(t)$.
{\bf c},
The spectrogram of the WWZ power of the Scaled light curve, in which a time-declining oscillation signal is evident.
{\bf d}, The light curve converted into phase space (i.e., the ``Stretched")
and the modeled $V(\theta)$.
{\bf e}, The spectrogram derived from the wavelet transform of the Stretched light curve.
{\bf f}, LSP method given power ($P_{\rm LSP}$) of the Scaled light curve and a shift time-averaged power spectrum ($P_W$) of the spectrogram in panel {\rm e} depict in gray line and gray dash-dotted line, respectively.
For panels {\bf e} and {\bf f}, the labels of the vertical axes are displayed on the right side, with the unit of frequency given in ``/rad"."
}
\label{fig:t-e periodicity}
\end{figure}
The GRB\,131122B triggered \textit{Fermi}/GBM, the Gamma-Ray Burst Monitor onboard the \textit{Fermi} spacecraft, on 22 November 2013 at 11:45:05 UT (denote $T_0$ hereafter) with $T_{90} = 23.04 \pm 0.81$ seconds \citep{2020ApJ...893...46V}.
Fig. \ref{fig:t-e periodicity}a displays the light curve of GRB\,131122B, with counts in the $8-900$ keV energy range, as recorded by the n4 detector of the \textit{Fermi}/GBM (quite a few other detectors have also recorded this event, but the signal to noise ratios are lower because of their much larger viewing angles or the higher background \citep{2020ApJ...893...46V} (see the Methods for the details).
The light curves exhibit a multi-pulse structure, with the peak times of these pulses showing progressively increasing time intervals.
The period enclosing the peaks identified by eye is  $t = T-T_0 \sim 10-28.3$ s.
The estimated pulse times are indicated by pink dashed lines, as shown in Fig. \ref{fig:t-e periodicity}a.
In Fig. \ref{fig:t-e periodicity}b we take a two smooth-broken power law function to approximate the median of the observed light curve in the time interval of $10-28.3$ s and then remove such a component to get the so-called ``Scaled" light curve (see the Methods for the details). The 2D plane contour plot of the weighted wavelet Z-transform (WWZ) power of the Scaled light curve is shown in Fig. \ref{fig:t-e periodicity}c,
which is distinguished by a time-decreasing frequency.
In particular, the Scaled light curve can be well fitted with $V(t)=17.30^{+0.90}_{-0.87} \cos \theta(t)$, where $\theta(t) =  39.64^{+0.92}_{-0.92}{\rm ln}(t-2.26^{+0.16}_{-0.17}) - 2.88^{+1.67}_{-1.72}$ (see the dashed red line in Fig. \ref{fig:t-e periodicity}b), corresponding to a rapidly decreasing frequency $f(t)=6.31^{+0.15}_{-0.15}(t-2.26^{+0.16}_{-0.17})^{-1}$ and revealing an oscillation signal with a period ($\propto f(t)^{-1}$) increasing almost linearly.
The oscillation signal consists of eight pulses, and the model predicting peak time at 10.63 s, 12.06 s, 13.72 s, 15.67 s, 17.95 s, 20.61 s, 23.73 s, and 27.38 s, respectively, corresponding to a period evolution from  1.27 s to 4.02 s within 16.75 seconds.
Based on the derived the evolution of periodicity, we converted the Scaled light curve into phase space by substituting $t$ with $\theta(t)$, and then we have $V(\theta) = 17.30 {\rm cos} \theta$.
The resulting ``Stretched'' light curve in phase space and its corresponding spectrogram are presented in Fig. \ref{fig:t-e periodicity}d and  Fig. \ref{fig:t-e periodicity}e, respectively.
A normalized power density spectra ($P_{\rm LSP}$) of Stretched light curve in the LSP method and a shifted time-averaged power spectrum ($P_W$) of spectrogram (Fig. \ref{fig:t-e periodicity}e) are present in Fig. \ref{fig:t-e periodicity}f, a signal with frequency $1/2\pi$ (/rad) is evident.
This procedure offers the advantage of intuitively highlighting potential periodic signals in the time series. 
This is an interesting QPO case selected among a few.
In the methods section, we employed a Gaussian Process to evaluate the significance of the observed time-evolving oscillation signature, in which the modeling have taking the uncertainties due to the transformation (stretching, etc.) applied to the original data into account (see Methods for details).

\begin{figure}[h]
\centering
\includegraphics[angle=0,scale=0.7]{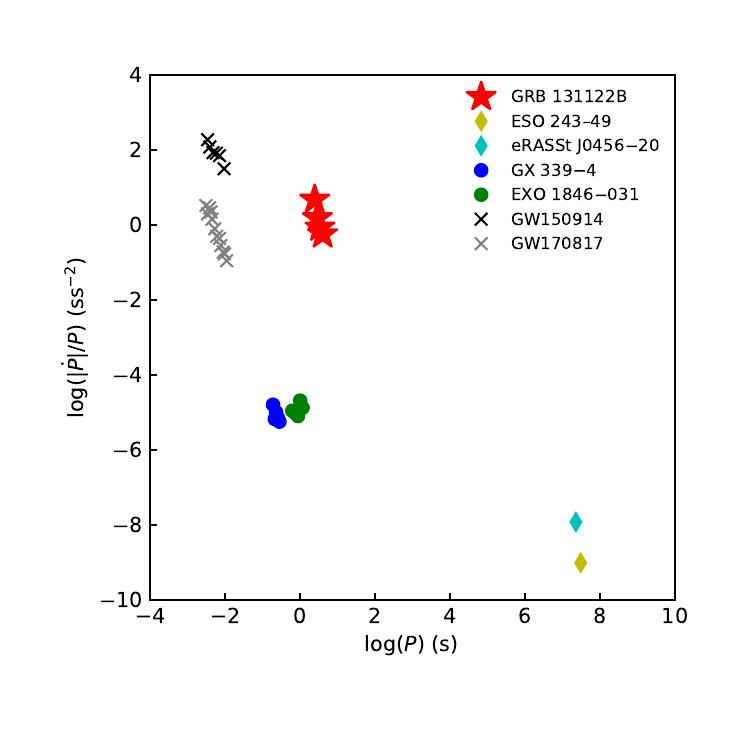}
\caption{\bf $|\dot{P}|/P - P$ diagram\rm.
The two GW events are adopted from \cite{2016PhRvL.116f1102A,2017PhRvL.119p1101A}, the potential pTDE candidates are from \cite{2020MNRAS.491.5682L} and \cite{2024A&A...683L..13L}, and the XRBs are taken from \cite{2024ApJ...971..148Z}. The eight pulses of GRB\,131122B are denoted with red stars. The time evolution of the oscillation signal displayed in GRB\,131122B is far quicker than any other electromagnetic QPO signals and it can even be comparable with that of GW 170817. Note that the GW signals have rapidly decreasing periods, in contrast with the electromagnetic one identified in this work.}
\label{fig:dotP}
\end{figure}

\section{The physical origin and discussions}
Previously, just  kilohertz QPO signatures may have been identified in the prompt emission light curves of two short-duration GRBs \citep{2023Natur.613..253C} and there is no evidence for time evolution of these QPO frequencies. Hence GRB\,131122B might be the first identified case of an event hosting an oscillation signal with an evolving period among the current GRB sample. 
Instead, the QPO frequency evolution has already been reported in some potential partial tidal disruption events (pTDEs) and X-ray binaries. For instance, hyperluminous X-ray source ESO 243-49 HLX-1
is characterized by an increase in the period from 300 days to over 700 days within seven years \citep{2014ApJ...793..105G}, while eRASSt J045650.3-203750 has a decreasing period, with an initial value of approximately 300 days and a reduction to 230 days in the last observations \citep{2024A&A...683L..13L}. These frequency evolutions are much more slower than what is found in GRB\,131122B. Somewhat quicker frequency evolution has been found in GX 339-4 and  EXO 1846-031 \citep{2024ApJ...971..148Z}. Nevertheless, such evolutions are still far slower than the signal of GRB\,131122B. In Fig. \ref{fig:dotP} we present a $|\dot{P}|/P- P$ diagram for the oscillation (or QPO) signals observed in these events, where $P$ is the period of the signal and $\dot{P}$ is its time derivative.
Clearly, the oscillation signal identified in GRB\,131122B has the quickest evolution among all the electromagnetic events of astrophysics.
Its evolution is so quick that can be comparable with that of the GW signal from the merger of double neutron stars (GW170817) \citep{2017PhRvL.119p1101A}, though slower than that of GW150914, the GW radiation from the merger of two BHs with tens of stellar masses \citep{2016PhRvL.116f1102A}. These facts strongly suggest that GRB\,131122B is related to a BH or a neutron star.


\begin{figure}[h]
\centering
\includegraphics[angle=0,scale=0.5]{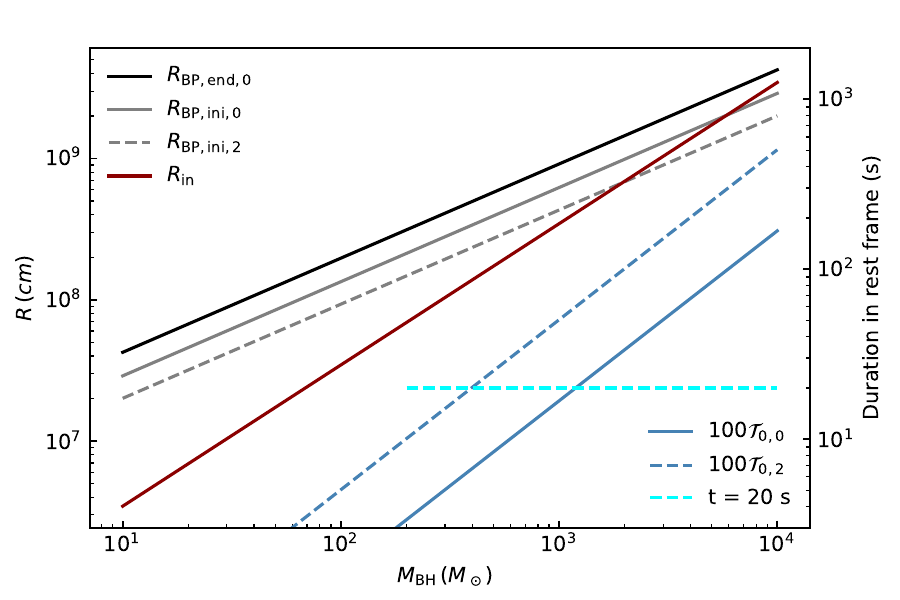}
\caption{\bf BH mass dependent parameters. \rm
The initial and the end precess radii ($R_{\rm BP, ini}$ and $R_{\rm BP, end}$) align with the time of the first and the last peak time, and the subscript indices 0 and 2 for $R_{\rm BP, ini}$ and $\mathcal{T}_{0}$ represent the scenarios at different redshifts.
}
\label{figEXT:phy_TDE}
\end{figure}

\subsection{Intermediate-mass BH induced tidal disruption event}
The QPO signals in XRBs are believed to originate from the instabilities of the corona-disk system \citep{2022A&A...662A.118M} or from the Lense-Thirring (LT) effect \citep{1918PhyZ...19..156L} induced precession of a truncated disk \citep{1998ApJ...492L..59S}.
Our current signal may be attributed to a precessing jet enslaved by an evolving disk.
Under the affection of the LT effect and the viscosity of materials, the inner region of a newly formed tilted disk surrounding a Kerr BH is expected to undergo rigid precession \citep{1975ApJ...195L..65B}.
The angular frequency of LT precession is $\Omega_{LT} = 2{\rm G}J/{\rm c}^2R_{\rm BP}^3$,
where $R_{\rm BP}$ is the characteristic precession radius,
and G and c are the gravitational constant and the speed of light, respectively.
The angular momentum $J$ is a function of the BH mass $M_{\rm BH}$ and its dimensionless spin $a$, written as $J = G M_{\rm BH}^2 a/{\rm c}$.
The precession period is expected to evolve rapidly during the early phase of disk formation. For a nearly constant $J$, $R_{\rm BP}\propto t^{1/3}$ is needed to account for the observed period evolution of the oscillation signal of GRB\,131122B. Interestingly, such a trend was indeed found in numerical simulations  \citep{2018MNRAS.474L..81L,2020arXiv200812381D} and the relevant duration is $t\leq 100 {\cal T}_0$, where ${\cal T}_0=2\pi R_{\rm g}(R_{\rm in}/R_{\rm g})^{3/2}/v_{k,in}$, $R_{\rm in}$ is the most inner stable orbit, $R_{\rm g}$ is the gravitational radius of the BH, and $v_{k,in} = GM_{\rm BH} /R_{\rm in}$ represents the Keplerian velocity at the innermost orbit.
The initial $R_{\rm BP,ini}$ (at the time of first peak) must be larger than $R_{\rm in}$. Assuming a BH spin parameter $a = 0.9$, for the cases of $z = 0$ and $z = 2$,  the required BH masses are no greater than $5860\,M_\odot$ and $1890\,M_\odot$, respectively.
The observed oscillation signal lasts for about 20 seconds, i.e., $100{\mathcal T}_0 \geq 20 /(1+z)$\,s.
Thus, the BH mass must be no less than $1190\,M_\odot$ and $390\,M_\odot$ for the cases of $z = 0$ and $z = 2$, respectively.
Without considering that GRB\,131122B has a very large redshift, e.g., $z > 9$, the BH mass should exceed $100\,M_\odot$. The relations among these parameters are presented in Figure~\ref{figEXT:phy_TDE}.
The supernova explosion can not directly produce such an intermediate-mass BH. To generate GRB\,131122B, the intermediate-mass BH should capture a star (i.e., GRB\,131122B is not a normal GRB, instead it resembles the tidal disruption event Swift J1644+57 \citep{2011Sci...333..199L}). The tidal radius of a star captured by such a BH can
be estimated by $R_{\rm t}\sim (7\times 10^{11}~{\rm cm})~(R_*/R_\odot)(M_*/M_\odot)^{-1/3}(M_{\rm BH}/ 10^3 M_\odot)^{1/3}$,
where $R_*$ and $M_*$ are the radius and mass of the captured star, respectively.
The bound part of material may create gamma-ray emission as it accretes onto the BH and most of the accretion is expected to be within a timescale of
$t_{\rm fall}\sim (9.46\times 10^3 \,{\rm s})\alpha_{0.1}^{-1} h^{-2} (R_p/R_t)^{3/2} (R_*/R_\odot)^{3/2} (M_*/M_\odot)^{1/2}$,
where $R_{\rm p}$ is the pericenter of the star's orbit, $\alpha$ is the viscosity parameter, $h$ is the ratio of disk height to radius \citep{1999ApJ...514..180U}.
By adopting $\alpha = 0.1$ and $h = 1$, to account for the $\sim30$ s duration of GRB,131122B, either a Sun-like star with $R_{\rm p} \sim R_{\rm t}/200$ or a white dwarf with $R_{\rm p} \sim R_{\rm t}/2$ is required. Therefore, a Sun-like star can be excluded, as the BH would penetrate it.
For the white dwarf scenario, one has $R_{\rm p}$ close to the maximum $R_{\rm BP}$ but a bit larger than it.
By assuming $M_{\rm BH} = 10^3 M_\odot$, we have $R_{\rm p} \sim  3.5\times 10^9 {\rm cm}$ and $R_{\rm t} \sim  7 \times 10^9 {\rm cm}$, respectively.

The best way to confirm the intermediate-mass BH induced TDE origin of GRB\,131122B is to accurately localize its position and then measure the afterglow emission as well as the properties of host galaxy.
This is however not achievable for GRB\,131122B since it has not been observed by other space telescopes in particular {\it Swift} and the localization is very poor, for which no follow up observations have been carried out. Nevertheless, the rate of the tidal disruption of a star by an intermediate-mass BH is much less frequent than that of the collapsar, which may account for the rarity of such signals in current GRB data.

\subsection{Could the oscillation signal be powered by an oblate magnetar?}\label{subsec:Physics_LT}

In the last section text, we have discussed a possible accretion disk precession origin of the observed oscillation signal. Here we examine the other possibility, in which the central engine is an oblate magnetar.
The ellipticity of the magnetar and its spin angular frequency determine the precession angular frequency $\Omega_p = \epsilon \Omega_m {\rm cos}(\theta)$, where ${\rm cos}(\theta) = \vec{\Omega}_m \cdot \vec{\epsilon} \sim 1 $ \citep{2015MNRAS.451..695Z}.
The observed period at the time of the first peak of GRB\,131122B is 1.27 seconds.
A magnetar born in a GRB typically has an initial rotation period of $P_0 =2\pi/\Omega_m \sim 1 - 10$ ms.
Therefore, the ellipticity of the magnetar is estimated as $\epsilon = P_0/ 1.27 s \sim 10^{-3} - 10^{-2}$.
To match the observed evolution of the oscillation signal requires $\epsilon \Omega_m \propto t^{-1}$.
The time dependence of $\Omega_m (t)$ exhibits three distinct regimes: $\Omega_m \propto t^0 $ before spin-down, $\Omega_m \propto t^{-1/4}$ for GW radiation dominated spin-down, and $\Omega_m \propto t^{-1/2}$ for magnetic dipole radiation dominated spin-down \citep{2001ApJ...552L..35Z}.
Here, we denote the ellipticity evolution of a newborn magnetar as $\epsilon \propto t^\xi $, and its power index is constrained to $ -1 \leq \xi \leq -1/2$.
In the supermassive neutron star model for some peculiar X-ray plateau followed by an abrupt flux drop, an $\epsilon$ high up to $\sim 10^{-2}$ was inferred, and a toroidal magnetic field as strong as $B_{\rm t}\sim 10^{17}(\epsilon/0.01)^{1/2}$\,G is needed to yield such a strong deformation of the magnetar \citep{2013PhRvD..88f7304F}. 
Furthermore, the evolution of current oscillation signal requires that the toroidal magnetic field inside the magnetar decays very quickly (i.e., $-1/2\leq \zeta \leq -1/4$ for $B_{\rm t}\propto t^\zeta $) to satisfy $ -1 \leq \xi \leq -1/2 $.
However, the independent evidence is currently lacking.

\subsection{Prospects}
All of the two short-duration GRBs with possible kHz QPO signals \citep{2023Natur.613..253C} and our current GRB\,131122B with a period-increasing oscillation signal have no identified counterparts in other bands, which hampers us to further understand these phenomena. Anyhow, the presence of such signals demonstrate that the high precision timing observations of GRBs can indeed open a new window to study the hiding central engine.
A future dedicated space gamma-ray observatory with a large effective area, a low background and a good angular resolution will be greatly helpful to identify the peculiar oscillation signals of GRBs and enable the catching of the counterparts in other wavelengths. The detection of such events together with in particular the associated gravitational wave radiation will be essential to better understand what happens to the hiding central engine (say, revealing the precession of the accretion disk) or even solve some top secrets in nuclear physics, for instance, the $\sim$ 3 kHz QPOs in short GRBs, likely relating to the post-merger gravitational-wave signal of double neutron star mergers, may effectively probe the presence of quark matter in the most massive neutron stars.


\begin{acknowledgments}
We acknowledge the use of the Fermi archive's public data. We thank the selfless discussions with Y.Z. Fan, B. Zhang, Q.Z Liu, X. Li, and S.P. Tang.
This work is supported by the Strategic Priority Research Program of the Chinese Academy of Sciences (grant no. XDB0550400),
We acknowledge the use of the BATSE, Fermi, and Swift archive's public data.
the National Key R\&D Program of China (2024YFA1611704),
the National Key Research and Development Program of China (2021YFA0718500), the National Natural Science Foundation of China (NSFC) under grant (Nos. 12473049, 12041301, 12121003), and the Postdoctoral Fellowship Program of CPSF under Grant Number GZC20241916.

\end{acknowledgments}



\appendix
\section{Extended data}\label{secA}

\subsection{Corner plot of median fit parameters}\label{secA1}

\begin{figure}[ht]
\centering
\includegraphics[angle=0,scale=0.3]{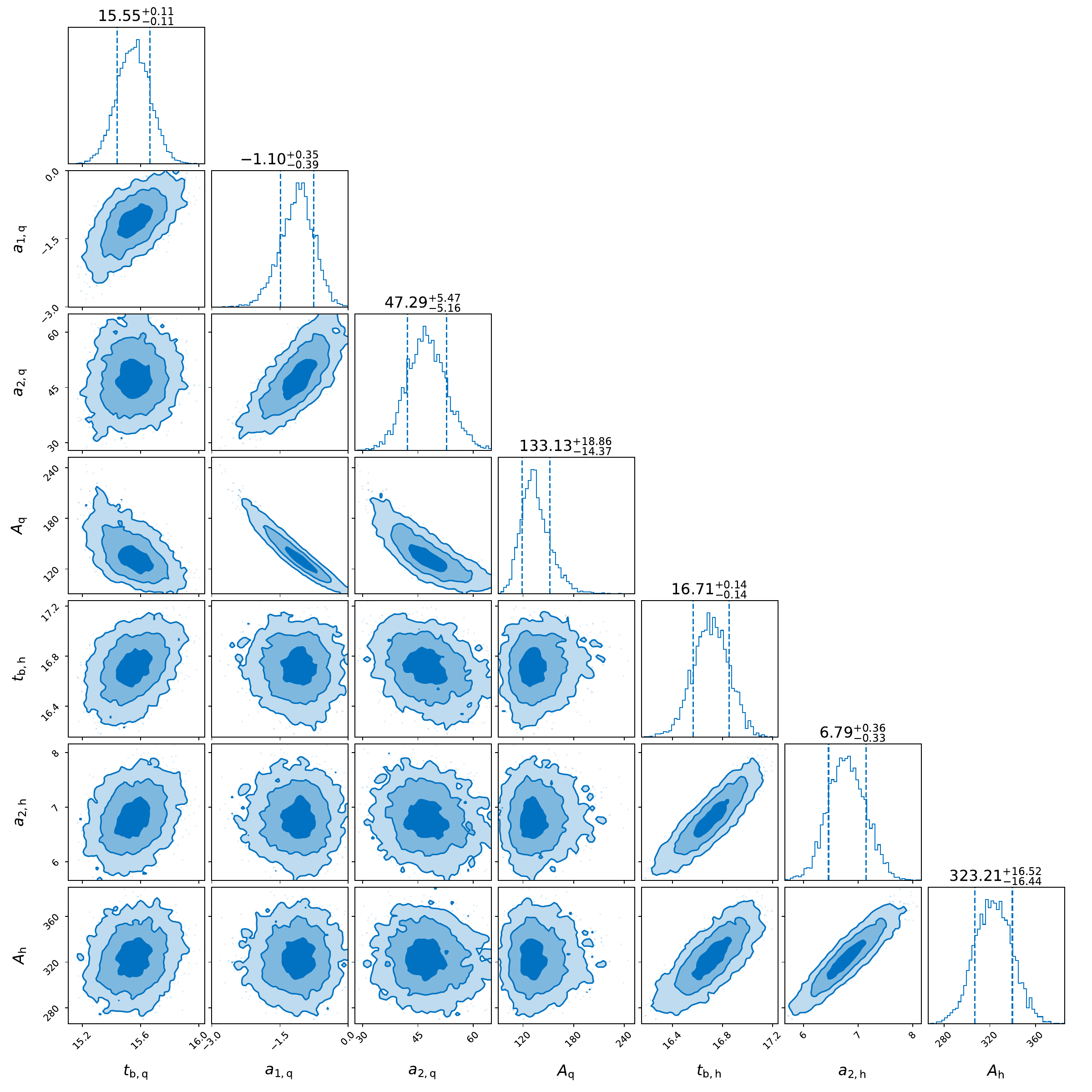}
\caption{\bf Corner plot of posterior parameter distribution in the median fit procedure\rm.
The performed results do not contain two fixed parameters $a_{\rm 1,h}$ and $w$.
}
\label{figEXT:corner_2SMBPL}
\end{figure}

\subsection{The unfixed $a_{1,h}$}\label{secA3}

\begin{figure}[ht]
\centering
\includegraphics[angle=0,scale=0.4]{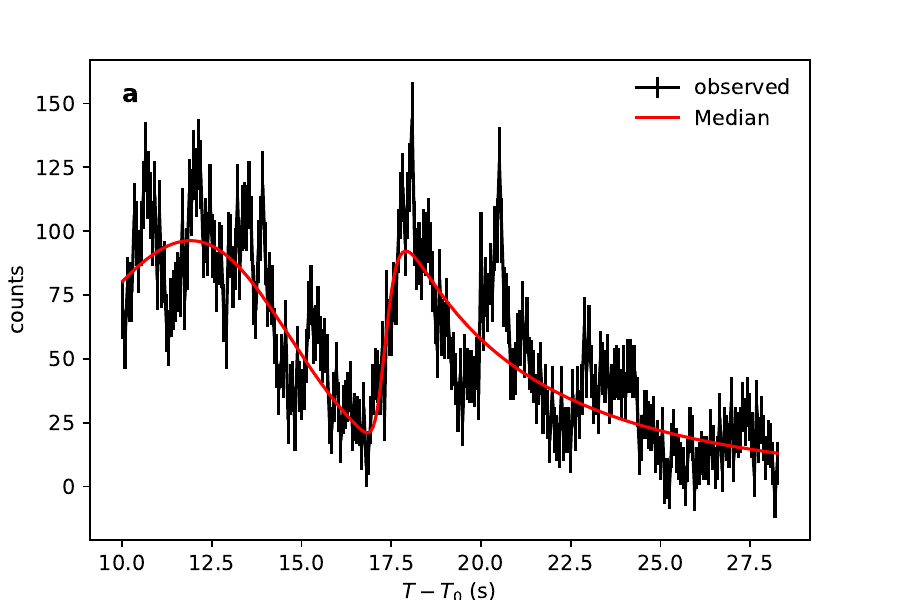}
\includegraphics[angle=0,scale=0.4]{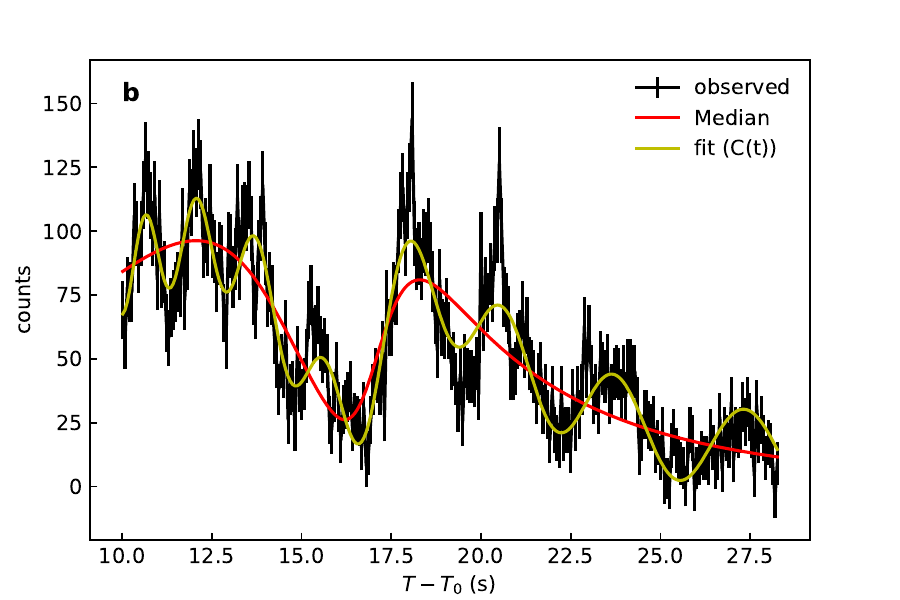}
\caption{\bf Demonstration for fitting procedure with unfixed $a_{\rm 1,h}$\rm.
{\bf a}, The Median is given by a fitting procedure using $f_b(t)$, which resulting in $a_{\rm 1,h} = -242.81^{+29.94}_{-30.59}$.
{\bf b}, The Median is given by a fitting procedure using $C(t)$, which resulting in $a_{\rm 1,h} = -80.41^{+10.66}_{-10.74}$.
}
\label{figEXT:test_a}
\end{figure}

\begin{figure}[ht]
\centering
\includegraphics[angle=0,scale=0.5]{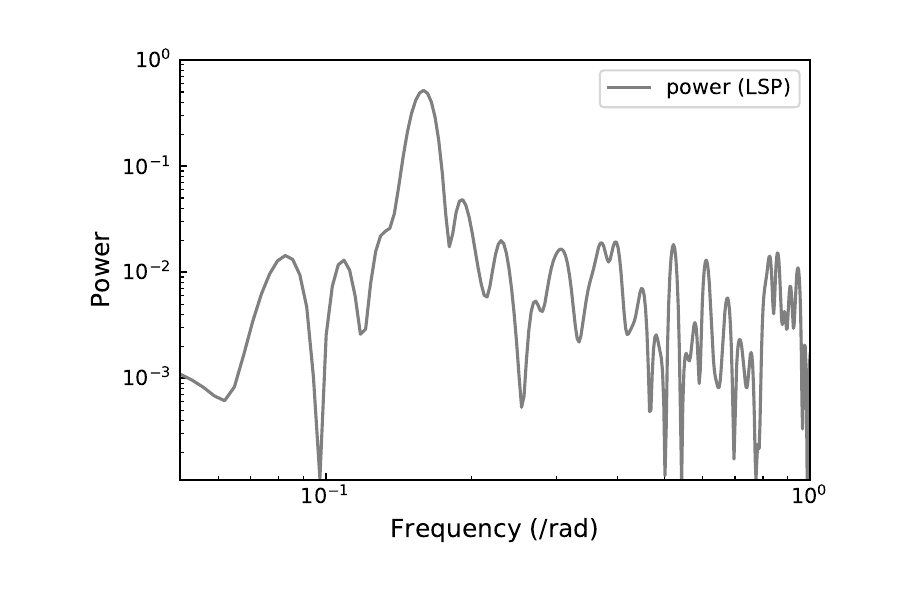}
\caption{\bf The periodogram given by LSP for fitting procedure using $C(t)$\rm.
}
\label{figEXT:trail_sig_a}
\end{figure}

In section \ref{subsec:normalization}, a fixed $a_{1,h}$ is adopted.
In an unfixed scenario, the value $a_{\rm 1,h} = -242.81^{+29.94}_{-30.59}$ is required, which results in a derived Median that aligns with the rapid rise of the fifth pulse (see Fig. \ref{figEXT:test_a}a).
Such a result deviates from our concept of ``median'', and should be regarded as an invalid result.
If we adopt a function  $C(t) = f_b(t) + V(t)$ to fit the observed light curve, the required $a_{\rm 1,h}$ decreases to $-80.41^{+10.66}_{-10.74}$.
In this case, the derived Median ($f_b(t)$) also somewhat aligns with the rapid rise of the fifth pulse (see Fig. \ref{figEXT:test_a}b).
The derived $f_b(t)$ used for periodic analysis shows negligible differences in both the shape of the LSP power and the significance of the characteristic signal, thus having no impact on our demonstration (see Fig. \ref{figEXT:trail_sig_a}).

\subsection{The comparison with other GRBs}\label{secA3}
\begin{figure}[ht]
\centering
\includegraphics[angle=0,scale=0.5]{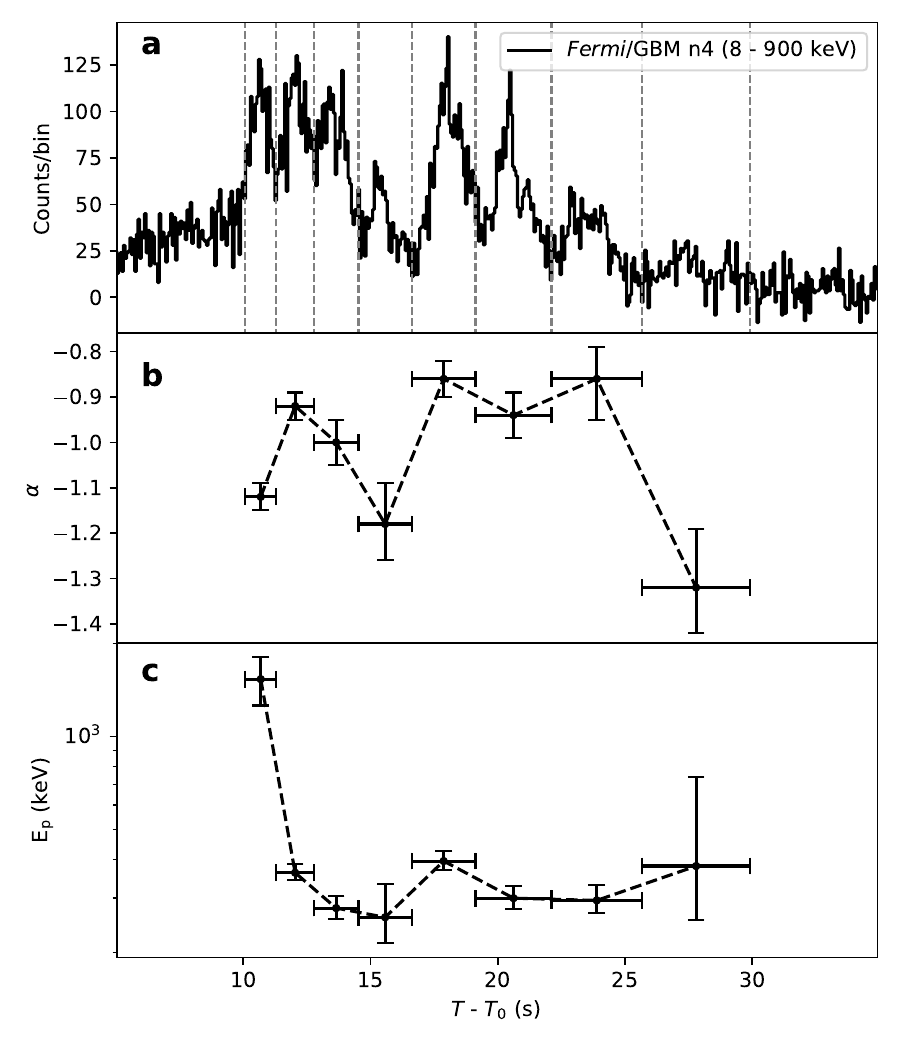}
\caption{\bf The time-resolved spectral analysis of GRB\,131122B\rm.
Based on the fitting result in the section \ref{subsec:evloution}, we selected eight time segments for the time-resolved spectra.
In panel {\bf a}, nine vertical gray dashed lines delineate the eight time segments.
The power law index $\alpha$ and the peak energy $E_p$ for the eight spectra are displayed in the panel {\bf b} and {\bf c}, respectively.
The peak energy, $E_p = (2+\alpha)E_c$, is defined for a $\nu F_\nu$ spectrum, characterized by flux tricking.}
\label{fig:Spectral_analysis}
\end{figure}

Based on the derived evolving phase angle of GRB\,131122B (see section \ref{subsec:evloution}), we selected eight time segments, each for a pulse bounded by the phase angles $(2n+1)\pi$ and $(2n-1)\pi$, for spectral analysis.
The specific time segments are 10.00 $- 11.32$ s, $11.32 - 12.86$ s, $12.86 - 14.66$ s, $14.66 - 16.76$ s, $16.76 - 19.23$ s, $19.23 - 22.11$ s, $22.11 - 25.48$ s, and $25.48 - 29.43$ s.
The TTE data from the n4 detector is adopted to generate the spectra 
and the background spectrum is estimated with the polynomial fit to the CSPEC data.
The data extraction was performed using the public {\tt GBM data tools}, while the forward folding of spectra and response matrices was carried out with {\tt PyXspec}.
The intrinsic photon spectrum was modeled as a cutoff power law $N(E)=A ({E}/{100\,{\rm keV}})^{\alpha}\exp(-{E}/{E_{\rm c}})$.
For Bayesian inference, we used the sampler {\tt pymultinest}, with the likelihood function being {PG-stat}.
The resulting time-resolved spectrum parameters are shown in Fig. \ref{fig:Spectral_analysis}.

\begin{figure}[ht]
\centering
\includegraphics[angle=0,scale=0.3]{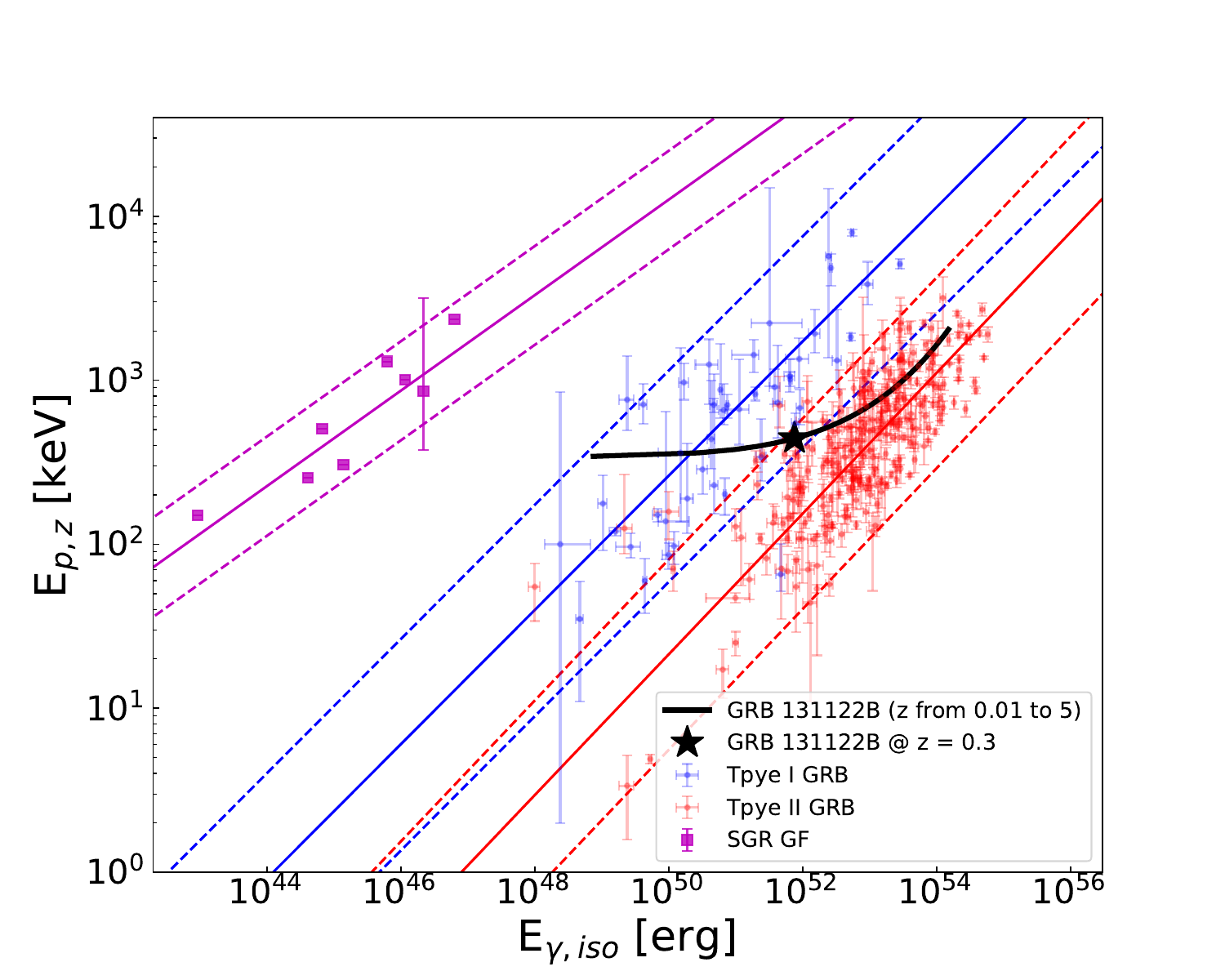}
\caption{\bf The $E_{\rm p,z}$ - $E_{\gamma,\rm iso}$ diagram\rm.
{The samples that type I and type II GRBs with known redshifts are adopted from \citep{2020MNRAS.492.1919M}}.
The trajectory of GRB\,131122B at different redshifts, represented by the black solid line, and the black star is for a redshift of 0.3.}
\label{fig:Amati}
\end{figure}

For the time-integrated spectrum, the energy peak $E_{\rm p}$, is about 342 keV, and the arrival flux, after considering the K-correction, is approximately 1.5 $\times 10^{-6}~{\rm  erg~cm^{-2}~s^{-1}}$.
A trajectory (black solid line) on the $E_{\rm p,z} - E_{\gamma,\rm iso}$ diagram \citep{2002A&A...390...81A} demonstrates the characteristic of GRB\,131122B local at  various redshifts ($0.01 \leq z \leq 5$).
It will be classified as a Type II GRB for $z\geq 0.3$. 

\clearpage

\bibliography{sn-bibliography}
\end{document}